\batchmode
\documentstyle[12pt] {article}
\def\zid{1\kern-0.36em\llap~1}

\newcommand{\beq}{\begin{equation}}
\newcommand{\ber}{\begin{eqnarray}}
\newcommand{\eeq}{\end{equation}}
\newcommand{\eer}{\end{eqnarray}}

\begin{document}

\begin{titlepage}
\rightline{SUNY BING 10/1/94}

\begin{center}
{\large \bf TESTS FOR TAU'S CHARGED-CURRENT
STRUCTURE}\\[2mm]
Charles A. Nelson\footnote{Electronic address: cnelson @
bingvmb.cc.binghamton.edu} \\
{\it Department of Physics\\State University of New York at
Binghamton\\
Binghamton, N.Y. 13902-6000}\\[5mm]
\end{center}

\begin{abstract}

The Lorentz structure of the tau lepton's charged-current can almost
be completely
\newline determined  by use of stage-two spin-correlation functions
for
the $\{\rho ^{-},\rho ^{+}\}$and $\{a_1^{-},a_1^{+}\}$ decay
modes.  It is
possible to test for a ``$(V-A)$ $ + $ something'' structure in the
${J^{Charged}}_{Lepton}$ current, so as to bound the scales
$\Lambda$ for ``new physics" such as arising from tau weak
magnetism,  weak
electricity, and/or second-class currents.
\end{abstract}

\begin{center}
{\bf Text}
\end{center}

Based on the assumption of a mixture of V and A couplings in the
$\tau$ charged-current, experiments at $e^- e^+$ colliders have been
setting limits
on the presence
of $(V+A)$ couplings in $\tau ^{-}\rightarrow A^{-}\nu _\tau $
decay for
$A=a_1, \rho, \pi, (l \bar{\nu_l})$.  The mixture of V and A
couplings can be
characterized by the value of the ``chirality parameter''
\beq
\xi _A\equiv
\frac{|g_L|^2-|g_R|^2}{|g_L|^2+|g_R|^2}=\frac{2Re\left(
v_Aa_A^{*}\right) }{%
|v_A|^2+|a_A|^2}.
\eeq
Note that $\xi _A=-\langle h_{\nu _\tau }\rangle $, twice the
negative of the $\nu _\tau $ helicity, in the special case for a spin-
one $A^{-
}$ particle of only $V$ and $A$
couplings and $m_\nu =0$. Using spin-correlations, the ARGUS
\cite{e1},
ALEPH  \cite{e2}, and CLEO \cite{e3} collaborations have
measured
$\xi _A$.
The current world average is $\xi
_A=1.002\pm 0.032$ \cite{e3,e4}.  So the leading contribution
in the tau's charged-current is $(V-A)$ to better than the $5\%$
level.

Therefore, the focus of this paper is on tests for ``something'' in a
{ \begin{quote}
(V-A) + ``something''
\end{quote} }
\noindent
structure in the tau's ${J^{Charged}}_{Lepton}$ current. This extra
contribution can
show up experimentally because of its interference with the $(V-A)$
part which,
we assume, arises as predicted by the standard lepton model.  More
precisely, the
idea is to search for ``additional structure'' due to additional Lorentz
couplings in
${J^{Charged}}_{Lepton}$ by generalizing the $\tau $ spin-
correlation function
$I(E_\rho ,E_{\bar B})$ by including the $\rho $ polarimetry
information
\cite{a2,c1} that is available from the $%
\rho^{ch}\rightarrow \pi^{ch}\pi ^o$ decay distribution \cite{C94}.
The
symbol $B=\rho
,\pi ,l$ . Since this adds on spin-correlation information from the
next
stage of decays in the decay sequence, we call such an energy-
angular
distribution a stage-two spin-correlation (S2SC) function. Similarly,
$a_1$
polarimetry information can be included from the  $\tau ^{-
}\rightarrow a_1^{-
}\nu \rightarrow \left( \pi ^{-}\pi
^{-}\pi ^{+}\right) \nu ,\left( \pi ^o\pi ^o\pi ^{-}\right) \nu $ decay
modes
\cite{C94a}.

The simplest useful S2SC is for the $CP$-symmetric decay sequence
$Z^o$, or
$%
\gamma ^{*}\rightarrow \tau ^{-}\tau ^{+}\rightarrow (\rho ^{-}\nu
_\tau
)(\rho ^{+}\bar \nu _\tau )$ followed by both $\rho ^{\mp
}\rightarrow \pi
^{\mp }\pi ^o$.
\ber
\text{I(E}_\rho \text{,E}_{\bar \rho }\text{,}\tilde \theta
_1\text{,}\tilde
\theta _{2\text{ }}\text{) = }|\text{T}\left( +-\right) |^2\rho
_{++}\bar
\rho _{--} +\text{ }|\text{T}\left( -+\right) |^2\rho _{--}\bar \rho
_{++} \nonumber  \\ +\text{ }|\text{T}\left( ++\right) |^2\rho
_{++}\bar \rho
_{++}
 +\text{ }|\text{T}\left( --\right) |^2\rho
_{--}\bar \rho _{--}
\eer
If we think in terms of probabilities, the quantum-mechanical
structure of this
expression is apparent,  since the $T(\lambda _{\tau
^{-}},\lambda _{\tau ^{+}})$ helicity amplitudes
describe the production of the $(\tau ^{-}\tau ^{+})$ pair via $Z^o$,
or $%
\gamma ^{*}\rightarrow \tau ^{-}\tau ^{+}$. For instance, in the 1st
term,
the factor  $|T(+,-)|^2=$``Probability to produce a $\tau ^{-}$ with
$%
\lambda _{\tau ^{-}}=\frac 12$ and a $\tau ^{+}$ with $\lambda
_{\tau
^{+}}=-\frac 12$ '' is multiplied by the product of the decay
probablity, $\rho _{++}$, for the positive helicity $\tau ^{-
}\rightarrow
\rho ^{-}\nu \rightarrow \left( \pi ^{-}\pi ^o\right) \nu $ times the
decay
probablity, $\bar \rho _{--}$, for the negative helicity $\tau
^{+}\rightarrow \rho ^{+}\bar \nu \rightarrow \left( \pi ^{+}\pi
^o\right)
\bar \nu $ .

The kinematic variables in $I_4$ are the usual ``spherical" ones
which naturally
appear in
the helicity formalism in describing such a decay sequence.  The 1st
stage of the
decay sequence $\tau
^{-},  \tau ^{+}\rightarrow (\rho ^{-}\nu _\tau )(\rho ^{+}\bar \nu
_\tau )$ is
described by the 3 variables $\theta _1^\tau,  \theta _2^\tau ,  \cos
\phi  $ where
$\phi$ is the opening $\angle $ between the two decay planes. These
are
equivalent
to the $Z^o$, or $\gamma ^{*}$
center-of-mass variables, $E_{\rho}, E_{\bar \rho }, \cos \psi $.
Here $\psi =$%
``opening $\angle $ between the $\rho ^{-}$ and $\rho
^{+}$momenta in the $%
Z/\gamma ^{*}$ cm''. When the Lorentz ``boost'' to one of the $\rho
$ rest frames
is directly from the
$Z/\gamma
^{*}$ cm frame, the 2nd stage of the decay sequence is described by
the
usual 2 spherical angles for the $\pi ^{ch}$ momentum
direction in that $\rho $ rest frame: $\tilde \theta _1, \tilde \phi _1$
for $\rho _1^{-
}\rightarrow \pi _1^{-}\pi _1^o$, and $\tilde \theta _2, \tilde \phi _2$
for $\rho
_2^{+}\rightarrow \pi _2^{+}\pi _2^o$. (Figures displaying these
variables are in
\cite{C94}.)

In (2), the composite decay density matrix elements are simply the
decay
probability for a $\tau _1^{-}$ with
helicity $\frac h2$ to decay $\tau ^{-}\rightarrow \rho ^{-}\nu
\rightarrow
\left( \pi ^{-}\pi ^o\right) \nu $ since
\beq
\frac{d\text{N}}{d\left( \cos \theta _1^\tau \right) d\left( \cos \tilde
\theta _1\right) }=\rho _{hh}\left( \theta _1^\tau ,\tilde \theta
_1\right)
\eeq

\noindent
and for the decay of the $\tau _2^{+}$ ,
\beq
\bar \rho _{hh}=\rho _{-h,-h}\left( {\text{subscripts}} \quad  1
\rightarrow 2, a
\rightarrow
b \right)
\eeq
We list the explicit formulas for $ \rho _{hh}\left( \theta _1^\tau
,\tilde \theta
_1\right)$, etc. in the appendix.

Note that the above $I_4$ spin-correlation function only depends on
4 of the
above 7 variables.  Refs. \cite{C94,C94a} give its generalization,
$I_7$, which
also depends on
$\cos{\phi}, \tilde \phi _1,$ and $\tilde \phi _2 $.  We use $I_4$ in
this paper
because it is less complicated and has a useful sensitivity level.
Sometimes $I_7$
is not significantly better in
regard to statistical errors^{\# 1}.

Ref. \cite{C94} only considered $\nu _L$ and $\bar \nu _R$
couplings.   It is
straightforward to include $\nu _R$ and $\bar \nu _L$ couplings in
$S2SC$
functions. The helicity amplitudes^{\#2} \hskip 1em   \rm  for \hskip
1em  $\tau^{-
}\rightarrow \rho
^{-}\nu _{L,R}$ for both $(V\mp A)$ couplings and
$%
m_\nu $ arbitrary are:
\newline for $\nu _L$ so $\lambda _\nu =-\frac 12$,%
\ber
A\left( 0,-\frac 12\right) & = & g_L\left(
\frac{E_\rho +q_\rho }{m_\rho }\right) \sqrt{m_\tau \left( E_\nu
+q_\rho
\right) } -g_R\left(
\frac{E_\rho -q_\rho }{m_\rho }\right) \sqrt{m_\tau \left( E_\nu -
q_\rho
\right) } \\ A\left( -1,-\frac 12\right) & = & g_L
\sqrt{2m_\tau \left( E_\nu +q_\rho \right) } -g_R\sqrt{2m_\tau \left(
E_\nu -q_\rho \right) }.
\eer
for $\nu _R$
so $\lambda _\nu =\frac 12$,%
\ber
A\left( 0,\frac 12\right) & = & -g_L\left(
\frac{E_\rho -q_\rho }{m_\rho }\right) \sqrt{m_\tau \left( E_\nu -
q_\rho
\right) }  +g_R\left(
\frac{E_\rho +q_\rho }{m_\rho }\right) \sqrt{m_\tau \left( E_\nu
+q_\rho
\right) } \\ A\left( 1,\frac 12\right) & = & -g_L
\sqrt{2m_\tau \left( E_\nu -q_\rho \right) } +g_R\sqrt{2m_\tau \left(
E_\nu +q_\rho \right) }
\eer
and $A\left( \pm 1, \mp \frac 12\right) =0$ by rotational invariance.
Notice that
$g_L,g_R$
denote the
`chirality' of the coupling and $\lambda _\nu =\mp \frac 12$ denote
the
handedness of $\nu _{L,R}$.

Historically in the study of the weak charged-current in muonic and
in
hadronic processes, it has been an important issue to determine the
``complete Lorentz structure'' directly from experiment in a model
independent manner. Here the $I_4$ and $I_7$ functions can be
used for this
purpose in investigating the $ \tau$ charged-current since these
functions
depend directly (see appendix) on the 4 helicity amplitudes for $\tau
^{-}\rightarrow \rho ^{-}\nu $ and on the 4 amplitudes for the $CP$-
conjugate
process. In this paper, for $I_4$ we report the associated ``ideal''
sensitivities.

We first consider the ``traditional'' couplings for  $\tau ^{-
}\rightarrow \rho
^{-}\nu $ which characterize the most general Lorentz coupling
$$
\rho _\mu ^{*}\bar u_{\nu _\tau }\left( p\right) \Gamma ^\mu u_\tau
\left(
k\right)
$$
where $k_\tau =q_\rho +p_\nu $. It is convenient to treat the vector
and
axial vector matrix elements separately. We introduce a parameter
$%
\Lambda =$ ``the scale of New Physics''. In effective field theory
this
is the scale at which new particle thresholds are expected to
occur.  In old-fashioned renormalization theory it is the scale at
which the calculational methods and/or the principles of
``renormalization''
breakdown, see for example \cite{th}. While some terms of the
following types do
occur as higher-order perturbative-corrections in the standard model,
such SM
contributions  are ``small'' versus the sensitivities of present tests in
$\tau$ physics
in the analogous cases of the $\tau$'s neutral-current and
electromagnetic-current
couplings, c.f.
\cite{d0}.  For charged-current couplings, the situation should be the
same.

In terms of the ``traditional'' tensorial and spin-zero couplings%
\ber
V_{\nu \tau }^\mu \equiv \langle \nu |v^\mu \left( 0\right) |\tau
\rangle
=\bar u_{\nu _\tau }\left( p\right) [g_V\gamma ^\mu
+\frac{f_M}{2\Lambda }\iota \sigma ^{\mu \nu }(k-p)_\nu
+\frac{g_{S^{-}}}{%
2\Lambda }(k-p)^\mu ]u_\tau \left( k\right) \\
A_{\nu \tau }^\mu \equiv \langle \nu |a^\mu \left( 0\right) |\tau
\rangle
=\bar u_{\nu _\tau }\left( p\right) [g_A\gamma ^\mu \gamma _5
+\frac{f_E}{2\Lambda }\iota \sigma ^{\mu \nu }(k-p)_\nu \gamma
_5+\frac{%
g_{P^{-}}}{2\Lambda }(k-p)^\mu \gamma _5]u_\tau \left( k\right)
\eer
\noindent
Notice that $\frac{f_M}{2\Lambda }=$ a ``tau weak magnetism''
type
coupling, and $\frac{f_E}{2\Lambda }=$ a ``tau weak electricity''
type
coupling. Both the scalar $g_{S^{-}}$ and pseudo-scalar  $g_{P^{-
}}$%
couplings do not contribute for  $\tau ^{-}\rightarrow \rho ^{-}\nu $
since $%
\rho _\mu ^{*}q^\mu =0$, nor for  $\tau ^{-}\rightarrow a_1^{-}\nu
$.

By Lorentz invariance, there is the equivalence theorem that for the
vector
current%
\ber
S\approx V+f_M, & T^{+}\approx -V+S^{-}
\eer
\noindent
and for the axial-vector current
\ber
P\approx -A+f_E, & T_5^{+}\approx A+P^{-}
\eer
where
\beq
\Gamma _V^\mu =g_V\gamma ^\mu +
\frac{f_M}{2\Lambda }\iota \sigma ^{\mu \nu }(k-p)_\nu   +
\frac{g_{S^{-}}}{2\Lambda }(k-p)^\mu +\frac{g_S}{2\Lambda
}(k+p)^\mu
+%
\frac{g_{T^{+}}}{2\Lambda }\iota \sigma ^{\mu \nu }(k+p)_\nu
\eeq
\beq
\Gamma _A^\mu =g_A\gamma ^\mu \gamma _5+
\frac{f_E}{2\Lambda }\iota \sigma ^{\mu \nu }(k-p)_\nu \gamma _5
+
\frac{g_{P^{-}}}{2\Lambda }(k-p)^\mu \gamma
_5+\frac{g_P}{2\Lambda }%
(k+p)^\mu \gamma _5  +\frac{g_{T_5^{+}}}{2\Lambda }\iota
\sigma ^{\mu \nu
}(k+p)_\nu \gamma _5
\eeq
The matrix elements of the divergences of these charged-currents are
\beq
(k-p)_\mu V^\mu =[g_V(m_\nu -m_\tau )
+
\frac{g_{S^{-}}}{2\Lambda }q^2+\frac{g_S}{2\Lambda }(m_\nu
^2-m_\tau ^2)
 +%
\frac{g_{T^{+}}}{2\Lambda }(q^2-[m_\tau -m_\nu ]^2)]\bar u_\nu
u_\tau
\eeq
\beq
(k-p)_\mu A^\mu =[g_A(m_\nu +m_\tau )
+
\frac{g_{P^{-}}}{2\Lambda }q^2+\frac{g_P}{2\Lambda }(m_\nu
^2-m_\tau ^2)
 +%
\frac{g_{T_5^{+}}}{2\Lambda }(q^2-m_\tau ^2+m_\nu ^2)]\bar
u_\nu \gamma
_5u_\tau
\eeq
Both the weak magnetism  $\frac{f_M}{2\Lambda }$ and the weak
electricty $%
\frac{f_E}{2\Lambda }$ terms are divergenceless. On the other
hand, since $%
q^2=m_\rho ^2$,  when $m_\nu =m_\tau $ there are non-vanishing
terms due to
the couplings $S^{-},T^{+},A,P^{-},T_5^{+}$.

Table 1 gives the limits on these additional couplings assuming a
``$(V-A)$ $+$%
something'' structure for the $\tau$ charged-current. Real coupling
constants
are assumed. Notice that at $M_Z$ the scale of $\Lambda \approx
$few $\
100GeV
$ can be probed; and at $10GeV$ or at $4GeV$ the scale of $1-
2TeV$ can be
probed.

All tables in this paper list only the ideal statistical errors \cite{c1},
and assume
respectively $10^7 Z^{o}$ events and $10^7$ (\tau^{-} \tau^{+} )$
pairs.
For the $\rho$ mode, we use B($\tau \rightarrow \rho \nu $) =
$24.6$\%. For the
$a_1$ mode we sum the charged plus neutral
pion $a_1$ final states so B($\tau \rightarrow {a_1}^{ch+neu} \nu
$) = $18$\%,
and use $m_{a_1} = 1.275GeV$.

The results in the tables simply follow by using (11-12) and from the
dependence
of the helicity amplitudes for $\tau ^{-}\rightarrow \rho ^{-}\nu $ on
the
presence of $(S\pm P)$ couplings with $m_\nu $ arbitrary:
\ber
A(0,-\frac 12) & =g_{S+P}(
\frac{m_\tau }{2\Lambda })\frac{2q_\rho }{m_\rho }\sqrt{m_\tau
(E_\rho
+q_\rho )}  +g_{S-P}(\frac{m_\tau }{2\Lambda })\frac{2q_\rho
}{m_\rho }%
\sqrt{m_\tau (E_\rho -q_\rho )}, \quad
A(-1,-\frac 12) & =0
\eer
and
\ber
A(0,\frac 12) & =g_{S+P}(
\frac{m_\tau }{2\Lambda })\frac{2q_\rho }{m_\rho }\sqrt{m_\tau
(E_\rho
-q_\rho )}  +g_{S-P}(\frac{m_\tau }{2\Lambda })\frac{2q_\rho
}{m_\rho }%
\sqrt{m_\tau (E_\rho +q_\rho )}, \quad
A(1,\frac 12) & =0
\eer

In compiling the entries in Table 1, we have adopted the idea of 1st
and 2nd
class currents \cite{sc1}. This is suggested by a 3rd-family
perspective of a possible ``$\tau \leftrightarrow \nu _\tau $
symmetry''
in the structure of the tau lepton currents. At the level of the masses,
this truly is a badly broken symmetry^{\# 3}.

But heeding the precedent historical successes of the SM in regard to
current-versus-mass symmetry distinctions, we believe that this
symmetry might
nevertheless be relevant to 3rd-family currents. Therefore, we
assume that the
effective charged-current ${J_{Lepton}}^{Charged}$ is Hermitian
and has such
an SU(2) symmetry, so that we can identify the $\nu _\tau $ and the
$\tau ^{-}$
spinors. Thereby, we obtain for the ``traditional couplings'' and real
form factors
that the ``Class I'' couplings are $V,A,f_M,P^{-}$, and that the
``Class
II'' couplings are  $f_E,S^{-}$ if we define  $J_{Lepton}^\mu =$
$J_I^\mu +$ $%
J_{II}^\mu $ where for $U=\exp (\iota \pi I_2)$%
$$
\begin{array}{cc}
(J_I^\mu )^{\dagger }=-UJ_I^\mu U^{-1} & First \\
(J_{II}^\mu )^{\dagger }=UJ_{II}^\mu U^{-1} & Second
\end{array}
Class
$$

This classification is particularly useful in considering the reality
structure of the charged-current \cite{sc2}. As show in Table 2 there
is a ``clash''
between the ``Class I and Class II'' structures and
the consequences of
time-reversal invariance. In particular, there are the useful theorems
that (a) ($\tau \leftrightarrow \nu _\tau $ symmetry) + ($T$
invariance) $%
\Longrightarrow $ Class II currents are absent, (b) ($\tau
\leftrightarrow
\nu _\tau $ symmetry) + (existence of $J_I^\mu $ and $J_{II}^\mu
$) $%
\Longrightarrow $ violation of $T$ invariance, and (c) (existence of
$%
J_{II}^\mu $) +  ($T$ invariance) $\Longrightarrow $($\tau
\leftrightarrow \nu _\tau $ symmetry) in $J_{Lepton}^\mu $ is
broken.

Table 3 shows the limits on such couplings assuming a pure-
imaginary
coupling constant. In the case of $(V-A)$ the limits on the $ \beta $'s
in Refs.
\cite{C94,C94a} cover this situation. Notice that the limits here are
in $ ( \Lambda
)^2$ and are $
\Lambda \sim$ few $10GeV$'s because, unlike for  Table 1, this is
not due to an
interference effect in the S2SC functions.

Besides the 3rd-family perspective of a possible $\tau \leftrightarrow
\nu _\tau $ symmetry, it is also instructive to consider ``additional
structure'' in the
$\tau$ charged-current from the
viewpoint of ``Chiral Combinations'' of the various
Lorentz couplings.  This is especially interesting because the $S\pm
P$ couplings
do not contribute to the transverse $\rho $ or $a_{1}$ transitions.
Tables 4 and 5
give the
limits on $\Lambda $ in the case of purely real and
imaginary coupling constants for the ``Chiral Couplings''.

Finally, as shown in Table 6,   the helicity amplitudes themselves
provide a simple framework for
characterizing a ``complete measurement'' of $\tau ^{-}\rightarrow
\rho
^{-}\nu $ decay and of $\tau ^{-}\rightarrow a_1^{-}\nu $ decay:  In
each case,
when only $\nu _L$ coupling's exist, there are only 2
amplitudes, so 3 measurements,  of $r_a,\beta _a,$and
$%
|A(0,-\frac 12)|$ via $\{\rho ^{-},B^{+}\}\mid _{B\neq \rho }$, will
provide
a ``complete measurement''. When $\nu _R$ coupling's also exist,
then
there are 2 more amplitudes, $A(0,\frac 12)$ and $A(1,\frac 12)$.
Then to
achieve
an ``almost'' complete measurement, 3 additional quantities must be
determined, e.g. by the $I_4$ S2SC function:  $r_a^R,\beta _a^R$
and $%
\lambda _R\equiv \frac{|A(0,\frac 12)|}{|A(0,-\frac 12)|}$. However,
to also
measure
the relative phase of the $\nu _L$ and $\nu _R$ amplitudes, $\beta
_a^o\equiv
\phi _o^{aR}-\phi _o^a$ or $\beta _a^1\equiv \phi _1^a-\phi _{-
1}^a$, requires,
e.g., the occurrence of a common final state which arises from both
$\nu _L$ and
$\nu _R$.

In conclusion,  $(\tau ^{-}\tau ^{+})$ spin correlations^{\# 4} \hskip
1em  \rm
with  \hskip 1em  $\rho$ and $a_1$ polarimetry observables can be
used to probe
for
``additional structure'' in the tau's charged-current. For example,
tau weak magnetism, $f_M(q^2)$,  and  tau weak
electricity, $f_E(q^2)$ ,can be probed to new physics scales of
\quad$\Lambda _{RealCoupling} \sim  1.2-1.5TeV$ \rm at $10$,  or
$4GeV$ and
\quad$\Lambda _{Imag.Coupling} \sim 28-34GeV$ \rm at $10,$ or
$4GeV$. \rm
By spin-correlation techniques the Lorentz stucture of the $\tau$
charge-current
can almost be completely determined from the $\{\rho ^{-},\rho
^{+}\}$and $%
\{a_1^{-},a_1^{+}\}$ modes.

\begin{center}
{\bf Acknowledgments}
\end{center}
For helpful discussions, we thank experimentalists and theorists at
Cornell,
DESY, Valencia, and at the Montreux workshop. This work was
partially
supported by U.S. Dept. of Energy Contract No. DE-FG 02-
96ER40291.
\begin{center}
{\bf Appendix: Useful formulas}
\end{center}
For a $\tau _1^{-}$ with
helicity $\frac h2$ to decay $\tau ^{-}\rightarrow \rho ^{-}\nu
\rightarrow
\left( \pi ^{-}\pi ^o\right) \nu $, the composite decay density matrix
elements are (
for only $\nu _L$
couplings)%
\ber
\rho _{hh}=
  \left( 1+h\cos \theta _1^\tau \right) \left[ \cos ^2\omega
_1\cos^2\tilde \theta
_1+\frac 12\sin ^2\omega _1\sin ^2\tilde \theta _1\right] \nonumber
\\
+ \frac{r_a^2}2\left( 1-h\cos \theta _1^\tau \right)
\left[ \sin ^2\omega_1\cos ^2\tilde \theta _1   +\frac 12\left( 1+\cos
^2\omega
_1\right) \sin^2\tilde \theta _1 \right]   \nonumber \\
 +h\frac{r_a}{\sqrt{2}}\cos \beta _a\sin \theta
_1^\tau \sin 2\omega _1\left[ \cos ^2\tilde \theta _1-\frac 12\sin
^2\tilde
\theta _1\right]
\eer
where $r_a$, $\beta_a$, and the Wigner rotation angle $\omega_1$
are defined
below. In the  $\tau ^{-}$ rest frame, the matrix element
for $\tau ^{-}\rightarrow \rho ^{-}\nu$ is
\beq
\langle \theta _1^\tau ,\phi _1^\tau ,\lambda _\rho ,\lambda _\nu
|\frac
12,\lambda _1\rangle =D_{\lambda _1,\mu }^{\frac 12*}(\phi
_1^\tau ,\theta
_1^\tau ,0)A\left( \lambda _\rho ,\lambda _\nu \right)
\eeq
where $\mu =\lambda _\rho -\lambda _\nu $.  Similarly, in the $\rho
^{-}$ rest
frame $\langle \tilde \theta _a,\tilde \phi _a|\lambda _\rho \rangle
=D_{\lambda
_\rho, 0}^1(\tilde \phi _a,\tilde \theta _a,0)c $ where $c$ is a
constant factor.  For
the $CP$-conjugate process, $\tau
^{+}\rightarrow \rho ^{+}\bar \nu \rightarrow \left( \pi ^{+}\pi
^o\right) \bar \nu $, in the $\tau ^{+}$ rest frame
\beq
\langle \theta _2^\tau ,\phi _2^\tau ,\lambda _{\bar \rho },\lambda
_{\bar
\nu }|\frac 12,\lambda _2\rangle =D_{\lambda _2,\bar \mu }^{\frac
12*}(\phi
_2^\tau ,\theta _2^\tau ,0)B\left( \lambda _{\bar \rho },\lambda
_{\bar \nu
}\right)
\eeq
with $\bar \mu =\lambda _{\bar \rho }-\lambda _{\bar \nu }$.  In the
$\rho ^{+}$
rest frame $\langle \tilde \theta _b,\tilde \phi _b|\lambda _{\bar \rho
}\rangle
=D_{\lambda _{\bar \rho }, 0 }^1(\tilde \phi _b,\tilde \theta _b,0)\bar
c $.

Assuming a L-handed $\nu _\tau $, $\tau ^{-}\rightarrow \rho ^{-
}\nu $
depends on the two amplitudes
\beq
A\left( -1,-\frac 12\right) =|A\left( -1,-\frac 12\right) |
\text{ e}^{\iota \phi _{-1}^a}, \qquad  A\left( 0,-\frac 12\right)
=|A\left(
0,-\frac 12\right) |
\text{ e}^{\iota \phi _0^a}.
\eeq
Similarly, assuming a R-handed $\bar \nu _\tau $%
, $\tau ^{+}\rightarrow \rho ^{+}\bar \nu $ depends on
\beq
B\left( 1,\frac 12\right) =|B\left( 1,\frac 12\right) |
\text{ e}^{\iota \phi _1^b}, \qquad B\left( 0,\frac 12\right) =|B\left(
0,\frac
12\right) |
\text{ e}^{\iota \phi _0^b}.
\eeq
By CP invariance, there are  2 tests for non-CKM-type leptonic CP
violation\cite{C94,C94a}: $\beta _a=\beta _b$ where $\beta _a=\phi
_{-1}^a-
\phi _0^a$,
$\beta
_b=\phi _1^b-\phi _0^b$, and $r_a=r_b $ where
\beq
r_a=\frac{|A\left( -1,-\frac 12\right) |}{|A\left( 0,-\frac 12\right) |},
\qquad r_b=%
\frac{|B\left( 1,\frac 12\right) |}{|B\left( 0,\frac 12\right) |}
\eeq
In the standard lepton model with a pure $(V-A)$ coupling, the
values of the
these polar parameters are $\beta _a=0,r_a=\frac{\sqrt{2}m_\rho
}{E_\rho
+q_\rho }\simeq \sqrt{2}m_\rho /m_\tau \simeq 0.613.$ In
$\rho_{hh}$, the
$\omega _1$ parameter is only a function of $\tilde \theta _1$
(i.e. of $%
E_\rho $) since
$$
\sin \omega _1=m_\rho \beta \gamma \sin \theta _1^\tau / p_1 ,
$$
$$
\cos \omega _1=\frac{M}{4m_\tau ^2p_1} \left( m_\tau ^2-m_\rho
^2 +\left[
m_\tau ^2+m_\rho ^2\right]
\beta \cos \theta _1^\tau \right)
$$
where $M=E_{cm}$, $\gamma =M/(2m_\tau )$.

For the $\tau ^{-}\rightarrow a_1^{-}\nu \rightarrow \left( \pi ^{-}\pi
^{-}\pi ^{+}\right) \nu ,\left( \pi ^o\pi ^o\pi ^{-}\right) \nu $ modes,
the
composite-decay-density matrix is given by (for only $\nu_L$
couplings)
 \ber
\rho _{hh}=
 \left( 1+h\cos \theta _1^\tau \right) \left[ \sin ^2\omega _1\cos
^2\tilde \theta _1
+ ( 1- \frac 12\sin ^2\omega _1 ) \sin ^2\tilde \theta _1\right]
\nonumber \\
+ \frac{r_a^2}2\left( 1-h\cos \theta _1^\tau \right)   \left[ \left(
1+\cos
^2\omega
_1\right) \cos ^2\tilde \theta _1  +\left( 1+\frac 12\sin ^2\omega
_1\right) \sin
^2\tilde \theta _1 \right]  \nonumber  \\
-h\frac{r_a}{\sqrt{2}}\cos \beta _a\sin \theta
_1^\tau \sin 2\omega _1\left[ \cos ^2\tilde \theta _1-\frac 12\sin
^2\tilde
\theta _1\right]
\eer
Here $\tilde \theta _1$ specifies the normal to the $\left( \pi ^{-}\pi
^{-}\pi ^{+}\right) $ decay triangle, instead of the $\pi ^{-}$
momentum
direction used for $\tau ^{-}\rightarrow \rho ^{-}\nu $.  The Dalitz
plot for $\left( \pi ^{-}\pi ^{-}\pi ^{+}\right) $ has been integrated
over
so that  \cite{bj} it is not necessary to separate the
form-factors for $a_1^{-} \rightarrow $ $\left( \pi ^{-}\pi ^{-}\pi
^{+}\right) $.

The resulting S2SC formulas are relatively simple for including $\nu
_R$ and
$\bar \nu
_L$ couplings,
\ber
I\left( E_\rho ,E_{\bar \rho },\tilde \theta _1,\tilde \theta _2\right)
\mid
_{\nu _R,\bar \nu _L}=I_4 +\left( \lambda _R\right) ^2I_4\left( \rho
\rightarrow
\rho ^R\right)
+\left( \bar \lambda _L\right) ^2I_4\left( \bar \rho \rightarrow \bar
\rho
^L\right) \nonumber  \\
+\left( \lambda _R\bar \lambda _L\right) ^2I_4\left( \rho \rightarrow
\rho
^R,\bar \rho \rightarrow \bar \rho ^L\right)
\eer
where
\ber
\lambda _R\equiv \frac{|A\left( 0,\frac 12\right) |}{|A\left(
0,-\frac 12\right) |}, \qquad  \bar \lambda _L\equiv \frac{|B\left( 0,-
\frac
12\right) |}{|B\left( 0,\frac 12\right) |}
\eer
give the moduli's of the $\nu _R$
and $\bar \nu _L$ amplitudes versus the standard amplitudes. The
corresponding
composite density matrices for $\tau \rightarrow \rho \nu $ with
$\nu _R$
and $\bar \nu _L$ final state particles are given by the substitution
rules:
\beq
\rho _{hh}^R=\rho _{-h,-h}\left( r_a\rightarrow r_a^R,\beta
_a\rightarrow
\beta _a^R\right) ,  \qquad
\bar \rho _{hh}^L=\bar \rho _{-h,-h}\left( r_b\rightarrow r_b^L,\beta
_b\rightarrow \beta _b^L\right)
\eeq
where the  $\nu _R$ and $\bar \nu _L$ moduli ratios and phase
differences
are defined by
\ber
r_a^R\equiv \frac{|A\left( 1,\frac 12\right) |}{%
|A\left( 0,\frac 12\right) |}, \qquad r_b^L\equiv \frac{|B\left( -1,-\frac
12\right) |}{|B\left( 0,-\frac 12\right) |},
\eer
and $\beta _a^R\equiv \phi _1^a-\phi
_0^{aR},\beta _b^L\equiv \phi _{-1}^b-\phi _0^{bL}$.

\begin{center}
{\bf Footnotes}
\end{center}
\begin{enumerate}
\item  For testing for $(V+A)$ versus  $(V-A)$, the use of the 3
additional variables $(\phi, \tilde{\phi_1}, \tilde{\phi_2} )$ in $I_7$
for $\{\rho
^{-},\rho ^{+}\}$ gives less than a $1\%$ improvement over $I_4$
at $M_Z$, $10
GeV$, or $4 GeV$.  If in addition the $\tau ^{-}$momentum
direction is
known via a SVX detector, there is only an $\sim 11\%$
improvement.  The same
numbers occur for $\{ {a_1}^{-
},{a_1}^{+} \}$.  In contrast, by using $I_4$, instead of the
simpler 2 variable $I\left( E_\rho ,E_{\bar \rho }\right) $ spin-
correlation
function, there is about a factor of 8 improvement at $M_Z$.
\item By $CP$ invariance, $B\left( \lambda _{\bar \rho },\lambda
_{\bar \nu
}\right) =A\left(-\lambda _{\bar \rho },-\lambda _{\bar \nu }\right)$,
which gives
the $\tau^{+}$ decay amplitudes with the Jacob-Wick phase
conventions.
By $CP\tilde T_{FS}$ invariance, B^{*}\left( \lambda _{\bar \rho
},\lambda
_{\bar \nu
}\right) =A\left( -\lambda _{\bar \rho },-\lambda _{\bar \nu }\right)$.
\item  Note $\frac{m_b}{m_t}\sim \frac 5{174}\sim 3\%$, and $%
\frac{m_\nu }{m_\tau }<\frac{23.8}{1777}\sim 1.4\%$ so this
symmetry is badly
broken in the masses for the 3rd family. However, for the other
leptons this
symmetry may be more strongly broken since
$\frac{m_{\nu
_e}}{m_e}<10^{-5}$, and $\frac{m _{\nu _\mu }}{m_\mu
}<0.15\%$ from the
current empirical bounds.  From phenomenological mass formulas,
e.g.
see \cite{har}, such as the GUT mass
formula, $\nu_{\tau}$:$\nu_{\mu}$:$\nu_{e}$ $\sim$
${m_t}^2$:${m_c}^2$:${m_u}^2$, the tau leptons are also the least
asymmetric
since then $\frac{m_{\nu
_\tau}}{m_\tau} \approx 10^{-8}$, $\frac{m_{\nu
_\mu}}{m_\mu} \approx 10^{-11}$, and $\frac{m_{\nu
_e}}{m_e} \approx 3\cdot 10^{-14}$ for the normalization
$m_{\nu_{\tau}}=20eV$.
\item  The tests in this paper make use of ($\tau^{-} \tau^{+}$) spin-
correlations
since it is assumed that the $e^{-}$ and $e^{+}$ colliding beams are
not
longitudinally-polarized.  It has recently been shown by Y.-S. Tsai
\cite{ch1,ch2} that in tau decays the sensitivities of tests for $CP$
violation, and
for other types of ``new physics'',  are substantially improved in
regard to both
systematic and statistical errors by the use of longitudinally-
polarized beams at the
($\tau^{-} \tau^{+}$) threshold.
\end{enumerate}

\begin{center}
{\bf Table Captions}
\end{center}
\quad Table 1: Limits on $\Lambda$ in $GeV$ for Real Coupling
Constants.  For
$V+A$ only, the entry is for $\xi_A$.

Table 2: ``Reality structure'' of $J^{\mu}_{Lepton}$ current's form
factors.

Table 3: Limits on $\Lambda$ in $GeV$ for Pure Imaginary
Coupling Constants.  For $V+A$ only, the entry is for $\xi_A$.

Table 4: ``Chiral Couplings'':  Limits on $\Lambda$ in $GeV$ for
Real Coupling Constants. For the $\rho$ and $a_1$ modes, the
$T^{+}+T_5^{+}$ coupling is equivalent to the $V-A$ coupling;
and $T^{+}-
T_5^{+}$ is equivalent to $V+A$.

Table 5: ``Chiral Couplings'':  Limits on $\Lambda$ in $GeV$ for
Pure Imaginary Coupling Constants. For the $\rho$ and $a_1$
modes, the
$T^{+}+T_5^{+}$ coupling is equivalent to the $V-A$ coupling;
and $T^{+}-
T_5^{+}$  is equivalent to the $V+A$.

Table 6: Elements of error matrix for limits on $\nu_R$ and
$\={\nu}_L$
couplings in terms of  the
helicity amplitudes for respectively $\tau \rightarrow \rho \nu$, and
$\tau
\rightarrow a_1 \nu$:

\newpage

\begin{table*}[h]
\setlength{\tabcolsep}{1.5pc}
\newlength{\digitwidth} \settowidth{\digitwidth}{\rm 0}
\catcode`?=\active \def?{\kern\digitwidth}
\caption{}
\label{tab1}
\begin{tabular*}{\textwidth}{@{}l@{\extracolsep{\fill}}rrrr}
\hline
                 & \multicolumn{2}{l}{$\lbrace \rho^{-}, \rho^{+}
\rbrace$ mode}
                 & \multicolumn{2}{l}{$\lbrace a_{1}^{-}, a_{1}^{+}
\rbrace$ mode}
\\
\cline{2-3} \cline{4-5}
                 & \multicolumn{1}{r}{At $M_Z$}
                 & \multicolumn{1}{r}{10, or 4 GeV}
                 & \multicolumn{1}{r}{At $M_Z$}
                 & \multicolumn{1}{r}{10, or 4 GeV}         \\
\hline
{\bf 1st Class Currents}  &                      &                   &
&               \\
$V+A$, for $\xi_A$        & $0.006$      & $0.0012$ & $0.010$ &
$0.0018$ \\
$f_M$, for $\Lambda$   & $214 GeV$ & $1,200$   &  $282$   &
$1,500$ \\
$S$                                   & $306 GeV$ &  $1,700$   &  $64$     &
$345$ \\
$T_5^{+}$                      & $506 GeV$ &  $2,800$   &  $371$   &
$2,000$ \\
{\bf 2nd Class Currents}&                      &                   &
&
\\
$f_E$, for $\Lambda$    & $214 GeV$ & $1,200$    &  $282$   &
$1,500$ \\
$P$                                   & $306 GeV$ &  $1,700$   &  $64$     &
$345$ \\
$T^{+}$                          & $506 GeV$ &  $2,800$   &  $371$   &
$2,000$ \\
\hline
\multicolumn{5}{@{}p{120mm}}{}
\end{tabular*}
\end{table*}

\begin{table*}[h]
\setlength{\tabcolsep}{1.5pc}
\newlength{\digitwidth} \settowidth{\digitwidth}{\rm 0}
\catcode`?=\active \def?{\kern\digitwidth}
\caption{}
\label{tab2}
\begin{tabular*}{\textwidth}{@{}l@{\extracolsep{\fill}}|rr|r}
\hline
 {\bf Form Factor:}             & Class I Current & Class II Current&
{\bf T
invariance} \\
\hline
$V,A,f_{M},P^{-}$           & Real parts          & Imaginary parts &
$Re\neq 0,
Im=0$  \\
$f_{E},S^{-}$                    & Imaginary parts & Real parts          &
$Re\neq 0,
Im=0$  \\
\hline
\end{tabular*}
\end{table*}

\begin{table*}[hbt]
\setlength{\tabcolsep}{1.5pc}
\newlength{\digitwidth} \settowidth{\digitwidth}{\rm 0}
\catcode`?=\active \def?{\kern\digitwidth}
\caption{}
\label{tab3}
\begin{tabular*}{\textwidth}{@{}l@{\extracolsep{\fill}}rrrr}
\hline
                 & \multicolumn{2}{l}{$\lbrace \rho^{-}, \rho^{+}
\rbrace$ mode}
                 & \multicolumn{2}{l}{$\lbrace a_{1}^{-}, a_{1}^{+}
\rbrace$ mode}
\\
\cline{2-3} \cline{4-5}
                 & \multicolumn{1}{r}{At $M_Z$}
                 & \multicolumn{1}{r}{10, or 4 GeV}
                 & \multicolumn{1}{r}{At $M_Z$}
                 & \multicolumn{1}{r}{10, or 4 GeV}         \\
\hline
{\bf 1st Class Currents:}     &                          &                  &
&
\\
$V+A$, for $\xi_A$             & $0.006$          & $0.0012$ & $0.010$
& $0.0018$
\\
$f_M$, for $(\Lambda)^2$ & $(12GeV)^2$ & $(28)^2$ &
$(15)^2$& $(34)^2$ \\
$S$                                         & $(14GeV)^2$ & $(33)^2$ & $(
6)^2$ & $(13)^2$
\\
$T_5^{+}$                            & $(22GeV)^2$ & $(50)^2$ &
$(18)^2$& $(42)^2$
\\
{\bf 2nd Class Currents:}     &                          &                  &
&
\\
$f_E$, for $(\Lambda)^2$   & $(12GeV)^2$ & $(28)^2$ &
$(15)^2$& $(34)^2$ \\
$P$                                         & $(14GeV)^2$ & $(33)^2$ & $(
6)^2$ & $(13)^2$
\\
$T^{+}$                                 & $(22GeV)^2$ & $(50)^2$ &
$(18)^2$& $(42)^2$
\\
\hline
\multicolumn{5}{@{}p{120mm}}{}
\end{tabular*}
\end{table*}

\begin{table*}[hbt]
\setlength{\tabcolsep}{1.5pc}
\newlength{\digitwidth} \settowidth{\digitwidth}{\rm 0}
\catcode`?=\active \def?{\kern\digitwidth}
\caption{}
\label{tab4}
\begin{tabular*}{\textwidth}{@{}l@{\extracolsep{\fill}}rrrr}
\hline
                 & \multicolumn{2}{l}{$\lbrace \rho^{-}, \rho^{+}
\rbrace$ mode}
                 & \multicolumn{2}{l}{$\lbrace a_{1}^{-}, a_{1}^{+}
\rbrace$ mode}
\\
\cline{2-3} \cline{4-5}
                 & \multicolumn{1}{r}{At $M_Z$}
                 & \multicolumn{1}{r}{10, or 4 GeV}
                 & \multicolumn{1}{r}{At $M_Z$}
                 & \multicolumn{1}{r}{10, or 4 GeV}         \\
\hline
$V+A$, for $\xi_A$                    & $0.006$         & $0.0012$ &
$0.010$ &
$0.0018$ \\
$S+P$, for $\Lambda$                & $310 GeV$   & $1,700$   &  $
64$     & $
350$ \\
$S-P$, for $(\Lambda)^2$          &$(11GeV)^2$& $(25)^2$ &
$(4)^2$  &
$(7)^2,(10)^2$ \\
$f_M+f_E$, for $\Lambda$      & $210 GeV$   & $1,200$   &
$280$   & $1,500$
\\
$f_M-f_E$, for $(\Lambda)^2$&$(9GeV)^2$  & $(20)^2$ &
$(10)^2$& $(24)^2$
\\
\hline
\multicolumn{5}{@{}p{120mm}}{}
\end{tabular*}
\end{table*}

\begin{table*}[hbt]
\setlength{\tabcolsep}{1.5pc}
\newlength{\digitwidth} \settowidth{\digitwidth}{\rm 0}
\catcode`?=\active \def?{\kern\digitwidth}
\caption{}
\label{tab5}
\begin{tabular*}{\textwidth}{@{}l@{\extracolsep{\fill}}rrrr}
\hline
                 & \multicolumn{2}{l}{$\lbrace \rho^{-}, \rho^{+}
\rbrace$ mode}
                 & \multicolumn{2}{l}{$\lbrace a_{1}^{-}, a_{1}^{+}
\rbrace$ mode}
\\
\cline{2-3} \cline{4-5}
                 & \multicolumn{1}{r}{At $M_Z$}
                 & \multicolumn{1}{r}{10, or 4 GeV}
                 & \multicolumn{1}{r}{At $M_Z$}
                 & \multicolumn{1}{r}{10, or 4 GeV}         \\
\hline
$V+A$, for $\xi_A$                    & $0.006$         & $0.0012$ &
$0.010$ &
$0.0018$ \\
$S+P$, for$(\Lambda)^2$          &$(11GeV)^2$& $(25)^2$ &
$(4)^2$  &
$(10)^2$ \\
$S-P$, for $(\Lambda)^2$          &$(11GeV)^2$& $(25)^2$ &
$(4)^2$  &
$(7)^2,(10)^2$ \\
$f_M+f_E$, for$(\Lambda)^2$&$(9GeV)^2$  & $(20)^2$ &
$(10)^2$& $(24)^2$
\\
$f_M-f_E$, for $(\Lambda)^2$&$(9GeV)^2$  & $(20)^2$ &
$(10)^2$& $(24)^2$
\\
\hline
\multicolumn{5}{@{}p{120mm}}{}
\end{tabular*}
\end{table*}

\begin{table*}[hbt]
\setlength{\tabcolsep}{1.5pc}
\newlength{\digitwidth} \settowidth{\digitwidth}{\rm 0}
\catcode`?=\active \def?{\kern\digitwidth}
\caption{}
\label{tab7}
\begin{tabular*}{\textwidth}{@{}l@{\extracolsep{\fill}}rrrr}
\hline
                 & \multicolumn{2}{l}{$\lbrace \rho^{-}, \rho^{+}
\rbrace$ mode}
                 & \multicolumn{2}{l}{$\lbrace a_{1}^{-}, a_{1}^{+}
\rbrace$ mode}
\\
\cline{2-3} \cline{4-5}
                 & \multicolumn{1}{r}{At $M_Z$}
                 & \multicolumn{1}{r}{10, or 4 GeV}
                 & \multicolumn{1}{r}{At $M_Z$}
                 & \multicolumn{1}{r}{10, or 4 GeV}         \\
\hline
{\bf Diagonal elements:}             &                     &
&
                       &    \\
$ a=\lambda_R$                           &(8\%)^2$    &$(4\%)^2$
&
$(18\%)^2$ &$(8\%)^2,(9\%)^2$\\
$ b=\lambda_Rr_a^R$                &$(8\%)^2$   &$(4\%)^2$
&
$(18\%)^2$ &$(8\%)^2,(9\%)^2$\\
$ c = $                                          &                     &
&
                      &  \\
$(\lambda_R)^2r_a^R
\cos{\beta^R}$
&$(13\%)^2$&$(6\%)^2,(10\%)^2$&$(41\%)^2$
                      &$(20\%)^2,(24\%)^2$ \\
{\bf Correlations:}                     &                     &
&
                      &    \\
$ \rho_{ab}$                               &$-0.75$       &$-0.77$
&$-0.95$
                      &$-0.96, -0.97$ \\
$ \rho_{ac}$                               &$-0.27$       &$-0.17,0.06$
&$-0.56$
                      &$0.029,0.019$ \\
$ \rho_{bc}$                               &$0.085$      &$.017,0.003$
& $0.04$
                      &$-0.41,-0.026$ \\
\hline
\multicolumn{5}{@{}p{120mm}}{}
\end{tabular*}
\end{table*}


\begin{thebibliography}{33}
\bibitem{e1} ARGUS collab., Phys. Lett. B250 (1990) 164; DESY
94-120; M. Schmidtler, {\em Third Workshop on Tau Lepton
Physics}, Montreux,
Switzerland, Sept. '94.
\bibitem{e2} ALEPH collab., Phys. Lett. B321 (1994) 168; J. Raab,
{\em Third
Workshop on Tau Lepton Physics}, Montreux,
Switzerland, Sept. '94.
\bibitem{e3}  R. Patterson, ICHEP94(Glasgow).
\bibitem{e4}M. Davier, {\em Third Workshop on Tau Lepton
Physics},
Montreux, Switzerland, Sept. '94; R. Stroynowski, ibid..
\bibitem{a2} J.H. Kuhn and F. Wagner, Nuc. Phys. B236 (1984) 16;
A. Rouge, Z. Phys C48 (1990) 75; M. Davier, L. Duflot, F.Le
Diberder, and A. Rouge, Phys. Lett. B306 (1993) 411; K. Hagiwara,
A.D. Martin, and D. Zeppenfeld, Phys. Lett., B235 (1990) 198; B.K.
Bullock, K. Hagiwara, and A.D. Martin, Nuc. Phys. B359 (1993)
499.
\bibitem{c1} C.A. Nelson, Phys. Rev. Lett. 62 (1989) 1347; Phys.
Rev. D40 (1989) 123; D41 (1990) 2327(E); Phys. Rev. D41 (1990)
2805; W.
Fetscher, Phys. Rev. D42 (1990) 1544; H. Thurn and H. Kolanoski,
Z. Phys. C60
(1993) 277.
\bibitem{C94} C.A. Nelson, H.S. Friedman, S.Goozovat, J.A. Klein,
L.R.
Kneller, W.J. Perry, and S.A. Ustin, Phys. Rev. D50 (1994)
4544; C.A. Nelson, SUNY BING 7/19/92; in {\em Proc. of
the Second Workshop on Tau Lepton Physics}, K.K. Gan (ed),
World Sci., Singapore, 1993.
\bibitem{C94a} C.A. Nelson, M. Kim, and H.-C. Yang, SUNY
BING 5/27/94; ICHEP94\#0100 (Glasgow); and a paper in
preparation.
\bibitem{th} G. 't Hooft, THU-94/15; S. Weinberg, in {\em
Unification of Elementary Forces and Gauge Theories}, D.B. Cline
and F.E. Mills(eds), Harwood Pub., London, 1978.
\bibitem{d0} S.M. Barr and W. Marciano, in {\em CP Violation},
C. Jarlskog(ed), World Sci., Singapore, 1989; W. Bernreuther, U.
Low, J.P. Ma,
and O. Nachtmann, Z. Phys. C43 (1989) 117; J. Bernabeu, N. Rius,
and A. Pich,
Phys. Lett. B257 (1991) 219; S. Goozovat and C.A. Nelson, Phys.
Rev. D44
(1991) 2818; J.A. Grifols and A. Mendez, Phys. Lett. B255
(1991) 611; and  R. Escribano and E. Masso, Phys. Lett. B301
(1993)
419; UAB-FT-317.
\bibitem{sc1} S. Weinberg, Phys. Rev. 112 (1958) 1375.
\bibitem{har} H. Harari, {\em Third Workshop on Tau Lepton
Physics},
Montreux, Switzerland, Sept. '94.
\bibitem{sc2} N. Cabibbo, Phys. Rev. Lett. 12 (1964) 137.
\bibitem{ch1} M. Perl, {\em Third Workshop on Tau Lepton
Physics}, Montreux,
Switzerland, Sept. '94.
\bibitem{ch2} Y.-S. Tsai, to appear in {\em Proc. of Workshop on a
Tau
Charm Factory in Era of CESR/B Factory}, SLAC, Aug. '94; Y.-S.
Tsai, SLAC-PUB-6685.
\bibitem{bj} S.M. Berman and M. Jacob, SLAC Report No. 43
(1965),
unpublished; Phys. Rev. 139 (1965) B1023.
\end{thebibliography}
\end{document}